\documentclass[osajnl,twocolumn,showpacs,superscriptaddress,10pt]{revtex4-1}
\usepackage{amsmath,amssymb}
\usepackage{graphicx}

\def\csch{\mathop{\rm csch}\nolimits}

\newcommand{\partialderiv}[3][]{\frac{\partial^{#1}#2}{\partial {#3}^{#1}}}
\newcommand{\dd}[2]{\frac{\partial {#1}}{\partial {#2}}}

\def\_#1{{\underline{#1}}}

\newcommand{\bp}{\begin{pmatrix}}
\newcommand{\ep}{\end{pmatrix}}

\newcommand{\be}{\begin{equation}}
\newcommand{\ee}{\end{equation}}

\newcommand{\sech} { {\rm sech} \hskip 0.01in}

\newcommand{\myRe}{\mathrm{Re}\,}
\newcommand{\myIm}{\mathrm{Im}\,}

\begin{document}

\title{Trade-off between Linewidth and Slip Rate in a Mode-Locked Laser Model}

\author{Richard O.\ Moore}
\affiliation{Dept.\ of Mathematical Sciences, NJIT, University Heights, Newark, NJ, 07102, USA}

\begin{abstract}
We demonstrate a trade-off between linewidth and loss-of-lock frequency in a mode-locked laser employing active feedback to control the carrier-envelope offset phase difference.  In frequency metrology applications, the linewidth translates directly to uncertainty in the measured frequency, while the impact of lock loss and recovery on the measured frequency is less well understood.  We reduce the dynamics to stochastic differential equations, specifically diffusion processes, and compare the linearized linewidth to the rate of lock loss determined by the mean time to exit calculated from large deviation theory.
\end{abstract}


\maketitle

\section{Introduction}

The advent of mode-locked lasers (MLLs) capable of generating frequency combs has revolutionized the field of frequency metrology by making available wide bands of stable lines from microwave to the visible regime and beyond~\cite{jun_ye_optical_2003,ruehl_advances_2012}.  These lines are determined by fixing the repetition rate and carrier-envelope offset (CEO) to a reference that currently resides in the microwave regime, although the possibility of a new standard in the optical regime has been proposed in the context of a new definition for the standard unit of time~\cite{jun_ye_optical_2003}.  The ability of MLLs to generate an entire octave of frequencies, either directly or using supercontinuum-generating fiber, allows self-referencing, where a low-frequency laser line is frequency doubled and interfered with the corresponding high-frequency line to provide a feedback signal that controls, for example, dispersion within the cavity.

The stability of these frequency sources is often limited by noise, stemming either from amplitude fluctuations of a pump laser or from spontaneous emission in the laser gain medium.  The impact of this noise is mitigated by an additional feedback in the form of a phase-locked loop~\cite{viterbi_principles_1966,cundiff_phase_2002} which, in combination with the self-referencing, is intended to restore the two primary degrees of freedom of the laser, the repetition rate and the CEO, to nominal values.
Standard measures of uncertainty of the frequency source are then obtained by taking moments of the output frequency distribution, the second of which gives the linewidth~\cite{ablowitz_noise-induced_2006,cundiff_femtosecond_????,haus_noise_1993,wahlstrand_quantum-limited_2008}.  This moment is well approximated by the linearized dynamics of the phase-locked loop about its set point.

Another source of uncertainty that is more difficult to characterize results from phase slips, where the phase of the voltage-controlled oscillator (VCO) in the phase-locked loop rotates by a full cycle.  At best, these phase slips introduce additional error in the output frequency of the laser; at worst, they lead to an unlocking of the feedback mechanism that may or may not be recoverable~\cite{hinkley_atomic_2013}.  These phase slips are assumed to result from exceedingly rare sequences of noise events, such that their probability of occurrence can be computed using large deviation theory.  Such a calculation provides important insight into whether measures to reduce linewidth inadvertently increase the likelihood of phase slips.

The present work analyzes a simple MLL model to explore the trade-off between output linewidth and mean time to phase slip.  Section~\ref{s:models} introduces the model considered and its reduction to a stochastic dynamical system.  Section~\ref{s:linear} analyzes the linearized model to compute the dependence of linewidth on physical parameters.  Section~\ref{s:LDT} applies recently developed techniques from large deviation theory to compute an effective loss-of-lock frequency, followed by concluding remarks in Section~\ref{s:discussion}.

\section{Mode-locked laser model with active feedback}
\label{s:models}

A phenomenological model of a Kerr lens MLL~\cite{cundiff_soliton_2005,donovan_rare_2007} is given by
\begin{align}
i\partialderiv{u}{z}&+\frac12\partialderiv[2]{u}{t}+|u|^2u = -b\cos(\omega t)u-ic_1u\nonumber\\
& +ic_2\partialderiv[2]{u}{t} +id_1|u|^2u -id_2|u|^4u + i\epsilon f(t,z),
\label{e:NLS_MLL}
\end{align}
where $u$ is the electric field envelope, $t$ is the retarded time variable normalized to the pulse width, and $z$ is the longitudinal variable normalized to a nonlinear length scale.  The left-hand side of Eqn.~\ref{e:NLS_MLL} thus accounts for chromatic dispersion and the Kerr nonlinearity.  The right-hand side includes a spatial phase modulation parametrized by $b$ and $\omega$ engineered to trap the pulse at $t=0$, saturable absorption and gain parametrized by $c_1$, $d_1$ and $d_2$, and optical filtering parametrized by $c_2$.  Physical considerations require that these constants be positive.  The noise process $f(t,z)$ is assumed to derive from spontaneous emission accompanying the stimulated emission on each pass through the gain medium, and assumed to be mean-zero Gaussian white noise, delta-correlated in $t$ and $z$.

To approximate the dynamics of optical pulses in this MLL model, we assume slow adiabatic changes in the pulse parameters with the following ansatz motivated by soliton solutions of Eqn.~\ref{e:NLS_MLL} with trivial right-hand side:
\begin{equation}
u_s(t,z) = A(z)\sech\left[A(z)\left(t-T(z)\right)\right]
e^{i\phi(z)+it\Omega(z)},
\label{e:Ansatz}
\end{equation}
where $A$, $T$, $\phi$ and $\Omega$ represent amplitude, position, phase and frequency.  The latter two parameters are actually offsets, measured relative to the underlying carrier signal of the MLL.  For PDE models that are variational (i.e., can be expressed as the stationarity condition of a Lagrangian functional), the Rayleigh-Ritz procedure~\cite{anderson_variational_2001} motivates a heuristic method for capturing the dynamics of the parameters in Eqn.~\ref{e:Ansatz} consisting of averaging over the Lagrangian density,
\begin{align}
L&[t,z,u(t,z),u_t(t,z),u_z(t,z)] = \nonumber\\
&\myIm(u_z^*u)-\frac12|u_t|^2+\frac12|u|^4+b\cos(\omega t)|u|^2,
\end{align}
to obtain a Lagrangian ${\mathcal L}[p(z)]=\int L[t,z,u_s,u_{st},u_{sz}]\,dt$ characterizing the dynamics on the four-dimensional manifold $p=(A, \Omega, T, \phi)^T$.
The presence of nonvariational terms in Eqn.~\ref{e:NLS_MLL}, however, requires the use of an extended version of this method~\cite{anderson_variational_2001,bale_variational_2008} that leads to the following inhomogeneous Euler-Lagrange equation for each parameter $p_j$, $j=1,\dots,4$:
\begin{align}
\dd{\mathcal L}{p_j}& - \dd{}{t}\dd{\mathcal L}{\dot{p_j}} =
2\,\myRe\int i\left[-c_1u_s+c_2u_{sxx}\right.\nonumber\\
& \left.+d_1|u_s|^2u_s-d_2|u_s|^4u_s+\epsilon f(x,t)\right]
\dd{u_s^*}{p_j}\,dx.
\end{align}

Carrying this out with the ansatz above gives a stochastic ordinary differential equation (SODE)
\begin{equation}
dU = F(U)\,dz + \epsilon \sigma(U)\,dW
\label{e:SODE}
\end{equation}
with
\begin{equation}
F = \bp -2c_1U_1 + (\frac43d_1-\frac23c_2)U_1^3-\frac{16}{15}d_2U_1^5 -2c_2U_1U_2^2\\
-\frac43c_2U_1^2U_2-\frac{\pi b\omega^2}{2U_1^3}\csch(\frac{\pi\omega}{2U_1})\sin(\omega U_3)\\
U_2\ep
\label{e:SODE_RHS}
\end{equation}
and
\begin{equation}
\sigma(U) = \bp \sqrt{U_1} & 0 & 0\\
0 & \sqrt{\frac{U_1}3} & 0\\
-\frac{U_3}{\sqrt{U_1}} & 0 & \sqrt{\frac{\pi^2}{12U_1^3}+\frac{U_3^2}{U_1}}\ep,
\end{equation}
where we have let $U=(A\; \Omega\; T)^T$.  Whether this SODE is to be interpreted in the sense of It\=o or Stratonovich depends on a more detailed derivation of Eqn.~\ref{e:NLS_MLL}; here, we assume the continuous noise process to be a limit of jump processes whose impact on the pulse parameters are appropriately derived from a prepoint analysis, and we therefore interpret Eqn.~\ref{e:SODE} using It\=o calculus.
The phase evolution is not included in the above dynamical system since it is slaved to the other three soliton parameters and does not influence the dynamics.  Its evolution is determined by
\begin{align}
d\phi &=\left[-T+\frac12(A^2-\Omega^2)\right.\nonumber\\
&\left.-\frac{\pi\omega b}{A^3}\cos(\omega T)\csch\left(\frac{\pi\omega}{2A}\right)\left(1+\frac{\pi\omega}{2A^2}\coth\left(\frac{\pi\omega}{2A}\right)\right)\right]dz\nonumber\\
&+\epsilon\left(\frac{\sqrt{12+\pi^2}}{6\sqrt{A}}\right)dW_4,
\label{e:phaseeqn}
\end{align}
where $W_4$ is a Wiener process independent from $dW$ above.  Clearly, $\phi(z)$ has a constant drift in $z$ at fixed values of $U$ unless the physical parameters are chosen to satisfy
\be
\frac12A_0^2=\frac{\pi\omega b}{A_0^3}\csch\left(\frac{\pi\omega}{2A_0}\right)\left(1+\frac{\pi\omega}{2A_0^2}\coth\left(\frac{\pi\omega}{2A_0}\right)\right).
\ee

\section{Linearized analysis}
\label{s:linear}

The fixed points of Eqn.~\ref{e:SODE} with $\epsilon=0$ are given by
\be
U_n = (A_0\; 0\; n\pi/\omega)^T
\ee
where $n$ is any integer and
\be
A_0^2 = \frac5{16d_2}\left[2d_1-c_2+
\sqrt{\left(2d_1-c_2\right)^2-\frac{96}5c_1d_2}\right],
\ee
provided
\be
\left(2d_1-c_2\right)^2>\frac{96}{5}c_1d_2.
\ee
The linearization about this fixed point takes the form $\dot{\delta U}=M\delta U$ with
\be
M = \bp M_{11} & 0 & 0\\
0 & M_{22} & M_{23}\\
0 & 1 & 0\ep
\label{e:linearization}
\ee
where $M_{11}=8c_1-\frac43(2d_1-c_2)A_0^2$, $M_{22}=-\frac43c_2A_0^2$ and  $M_{23} = (-1)^{n+1}\frac{\pi b\omega^3}{2A_0^3}\csch(\frac{\pi\omega}{2A_0})$.
Recalling that all of the physical constants are positive, the eigenvalues of $M$ indicate that the fixed points with $n$ even are stable, while those with $n$ odd are unstable.  The stable fixed points are nodes if
\be
\left(\frac43c_2A_0^2\right)^2>\frac{2\pi b\omega^3}{A_0^3}\csch\left(\frac{\pi\omega}{2A_0}\right);
\ee
otherwise they are spirals.  Without loss of generality, we take the $n=0$ fixed point as the desired operating condition of the MLL.
The system also approaches stationarity for $\Omega=0$ as $A$ approaches zero for arbitrary $T$; this line of stable fixed points simply reflects the stability of the trivial operational state of the MLL.

From the linearization in Eqn.~\ref{e:linearization} one can derive the spectral density of $U$ and therefore the variances of $A$, $\Omega$ and $T$ in steady state.  They are the diagonal elements~\cite{viterbi_principles_1966} in
\be
\Sigma_U = \frac{\epsilon^2}{2\pi}\int ds \left[(is I-M)^{-1}\sigma\right]\left[(is I-M)^{-1}\sigma\right]^{\dag}.
\ee
Specifically,
\begin{align}
&\sigma_A^2 = \frac{\epsilon^2A_0}{2|M_{11}|},\\
&\sigma_\phi^2 = \frac{\sqrt{2}A_0\epsilon^2}{6\chi}\left(1+\frac{\pi^2|M_{23}|}{4A_0^4}\right),\\
&\sigma_T^2 = \frac{\sqrt{2}\pi^2\epsilon^2}{24A_0^3\chi}\left(1+\frac{M_{22}^2}{|M_{23}|}\right),
\end{align}
where
\begin{align}
\chi = &\left(\sqrt{M_{22}^2+2M_{23}+\sqrt{M_{22}^4+4M_{22}^2M_{23}}}\right.\nonumber\\
 &+ \left.\sqrt{M_{22}^2+2M_{23}-\sqrt{M_{22}^4+4M_{22}^2M_{23}}}\right).
\end{align}

\section{Exit problem}
\label{s:LDT}

Equally important to the stable and effective operation of the laser is its likelihood to skip phase cycles, i.e., to incur a ``phase slip''.  The feedback mechanism in a mode-locked laser with carrier-envelope phase control is a phase-locked loop, where the loop cannot distinguish between phases $\phi+2\pi n$ with $n\in\mathbb{Z}$.  A noise-induced $2\pi$ phase rotation, at best, adds to variability of the laser frequency (or to the counter, if the laser is being used for timekeeping).  Depending on the feedback mechanism, a loss of lock can lead to counting errors or failure~\cite{hinkley_atomic_2013}.

In the model under consideration, we associate a phase slip with an exit event from the basin of attraction of $(A_0,0,0)^T$ under the dynamics of Eqn.~\ref{e:SODE}.  This is not strictly correct, since the phase $\phi$ can undergo a $2\pi$ rotation due to either (or both) of the forcing terms in Eqn.~\ref{e:phaseeqn}; however, we restrict ourselves to physical parameters for which escape of $U$ to a different stable fixed point is a more likely slip mechanism than the simple random walk of $\phi$.
To infer a phase slip {\em rate}, we seek to compute the mean exit time, which can be obtained from exit statistics on suitably large time intervals.  Given the small probability of exit, this quantity is dominated by the minimizer of the quasi-potential, i.e., the minimizer over all times of the Wentzell-Freidlin (W-F) action~\cite{freidlin_random_2012},
\begin{align}
S_{\infty}&=\inf_{\psi} \frac12\int_0^{\infty}(\dot{\psi}-F(\psi))^T(\sigma(\psi)\sigma(\psi)^T)^{-1}(\dot{\psi}-F(\psi))\,dt,
\label{e:WentzellFreidlin}
\end{align}
where the infimum is taken over all absolutely continuous paths $\psi$ that start at $(A_0\; 0\; 0)^T$ and end at either of the nearest saddles $(A_0\; 0\; \pm\pi/\omega)^T$.  The minimizing path $U_\mathrm{opt}$ is the most likely path leading to exit of the basin of attraction of the stable fixed point.  By symmetry, we need only consider the path exiting through $(A_0\; 0\; \pi/\omega)^T$ and double the resulting probability (or inferred rate).  As $\epsilon\rightarrow 0$, the expected first exit distance $z_{\mathcal D}$ from domain ${\mathcal D}$, taken here to be the basin of attraction of the desired operating point, scales according to
\be
\ln z_{\mathcal D}  \sim S_{\infty}/\epsilon^2.
\label{e:meanexitrate}
\ee
The logarithmic dependence in this asymptotic relationship between mean first exit distance and the W-F action functional allows for an arbitrary multiplicative constant in $z_{\mathcal D}$ that is generally not calculable without using sampling, although in specific cases this prefactor can be approximated using WKB analysis~\cite{bobrovsky_singular_1982}.  The mean first exit time is then related to the mean first exit distance through the group velocity $v_g$, providing an effective phase slip rate of $\nu_{\mathrm{slip}}=v_g/z_{\mathcal D}$.

\section{Computing $S_\infty$}
\label{s:sinf}

To find $S_\infty$ and the associated highest-likelihood path dominating the mean first exit time computation, we use a rescaling of the functional in Eqn.~\ref{e:WentzellFreidlin} that allows the minimization to be performed with respect to (finite) arclength of the path rather than (infinite) time taken along the path.  This is the basis of the geometric minimimum action method~\cite{heymann_geometric_2008}, where the minimizer can be found using a straightforward relaxation or gradient descent algorithm to solve the two-point boundary value problem obtained by applying the calculus of variations.  We refer the reader to Ref.~\cite{heymann_geometric_2008} rather than include the details here.

The figures below were generated using physical parameters of $c_1=c_2=d_2=1$, $d_1=3$, and $b=0.85$ (chosen to set the first term on the right-hand side of Eqn.~\ref{e:phaseeqn} to zero), giving an equilibrium amplitude of $A_0=1.5$.  The optimal path connecting the two fixed points $(A_0\; 0\; 0)^T$ and $(A_0\; 0\; 2)^T$ was discretized into 2048 arclength segments, with the derivatives computed using finite differences and a relaxation method with nonadaptive artificial time step of $10^{-4}$.

\begin{figure}[htbp]
\centerline{%
\includegraphics[width=1\columnwidth]{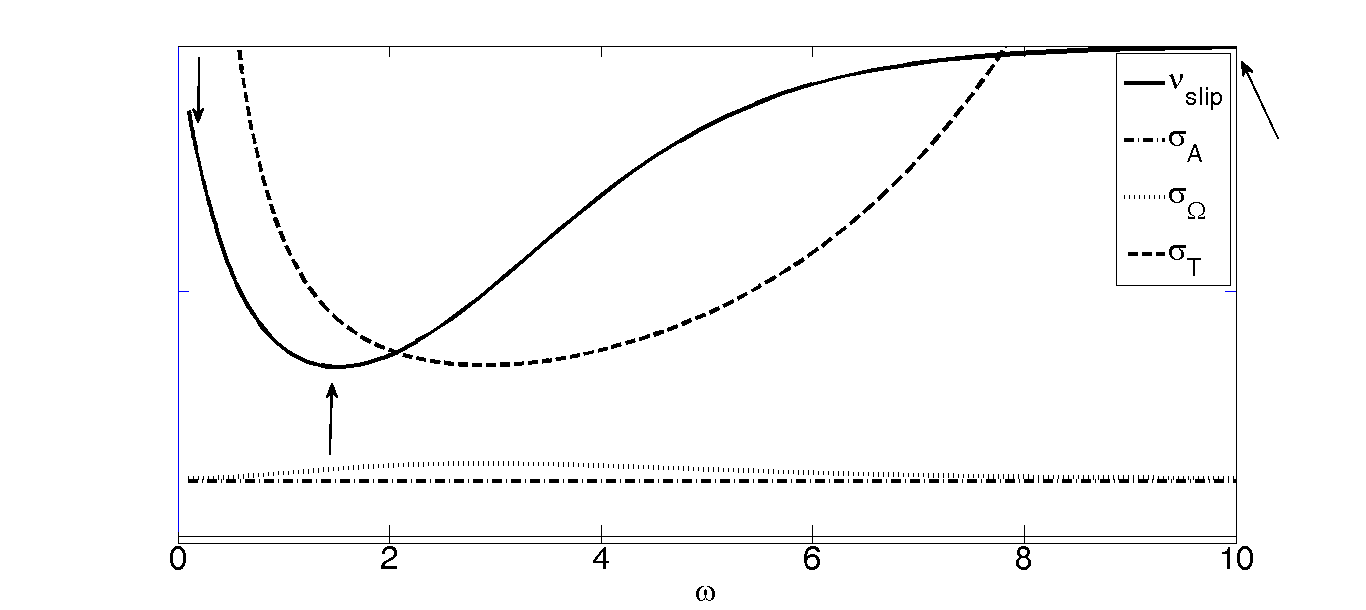}}
\caption{Measures of uncertainty versus control parameter $\omega$.  Plotted are phase slip rate (solid) and standard deviations of amplitude $A$ (dashed), frequency $\Omega$ (dotted), and timing $T$ (dashed-dotted).  Arrows indicate values of $\omega$ used to generate Figs.~\ref{f:omega1} through~\ref{f:omega10}.}
\label{f:tradeoff}
\end{figure}

Figure~\ref{f:tradeoff} plots the phase slip rate and long-time standard deviations of each parameter against control parameter $\omega$, which parametrizes the width of the active feedback potential engineered to restore the optical pulses to $T=0$.  This figure illustrates the primary result of this article, the fact that optimizing the feedback parameters for minimum linewidth does not yield optimal control parameters ($\omega$, in this case) for minimizing phase slip rate. Which of these criteria is more important depends on the impact of a phase slip on long-term operation of the laser.  If phase slips cause catastrophic failure of the feedback mechanism, then it is obviously most important to minimize these events; on the other hand, if the feedback mechanism quickly resets after a phase slip, a choice of linewidth-optimal parameters might be more appropriate.  If the impact of phase slips is not fully understood for a given feedback mechanism, Eqn.~\ref{e:meanexitrate} provides an estimate for how often one should expect such phase slips to occur, for comparison with experimental results.

\begin{figure}[htbp]
\centerline{%
\includegraphics[width=\columnwidth]{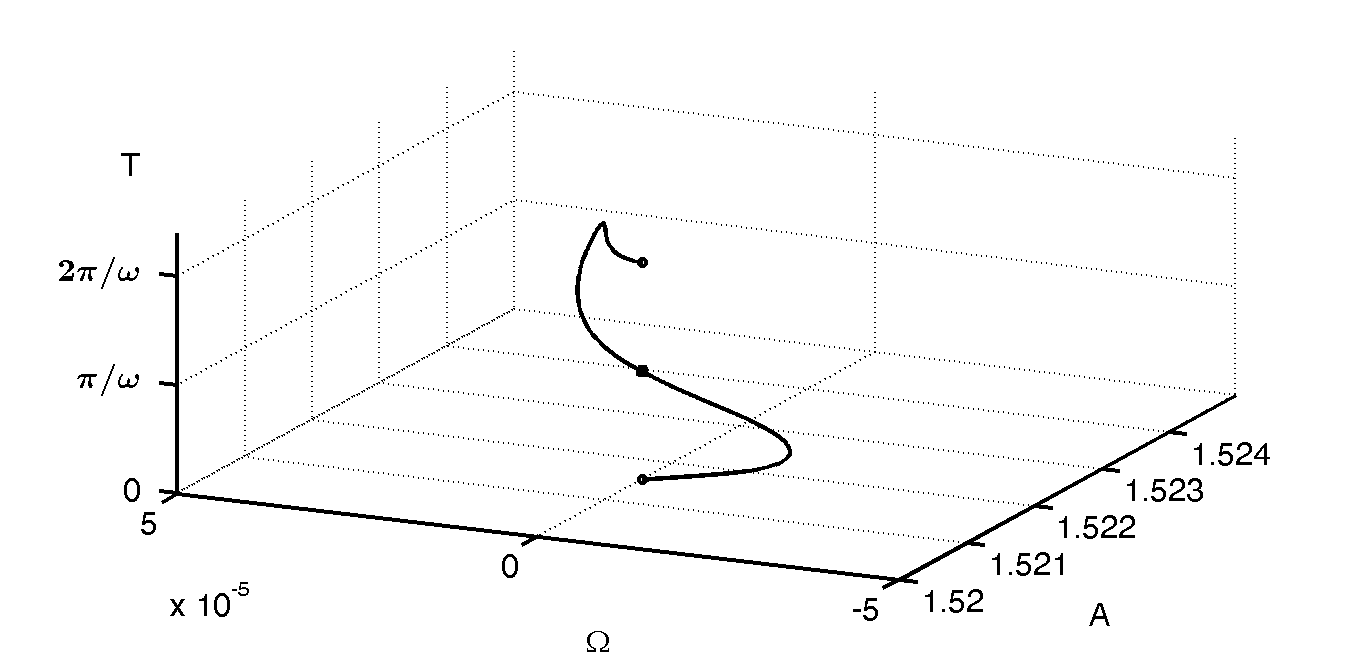}}
\caption{Optimal path for $\omega=0.1$.  Fixed points along path are indicated by circles (stable fixed points) or squares (saddles).}
\label{f:omega1}
\end{figure}

\begin{figure}[htbp]
\centerline{%
\includegraphics[width=\columnwidth]{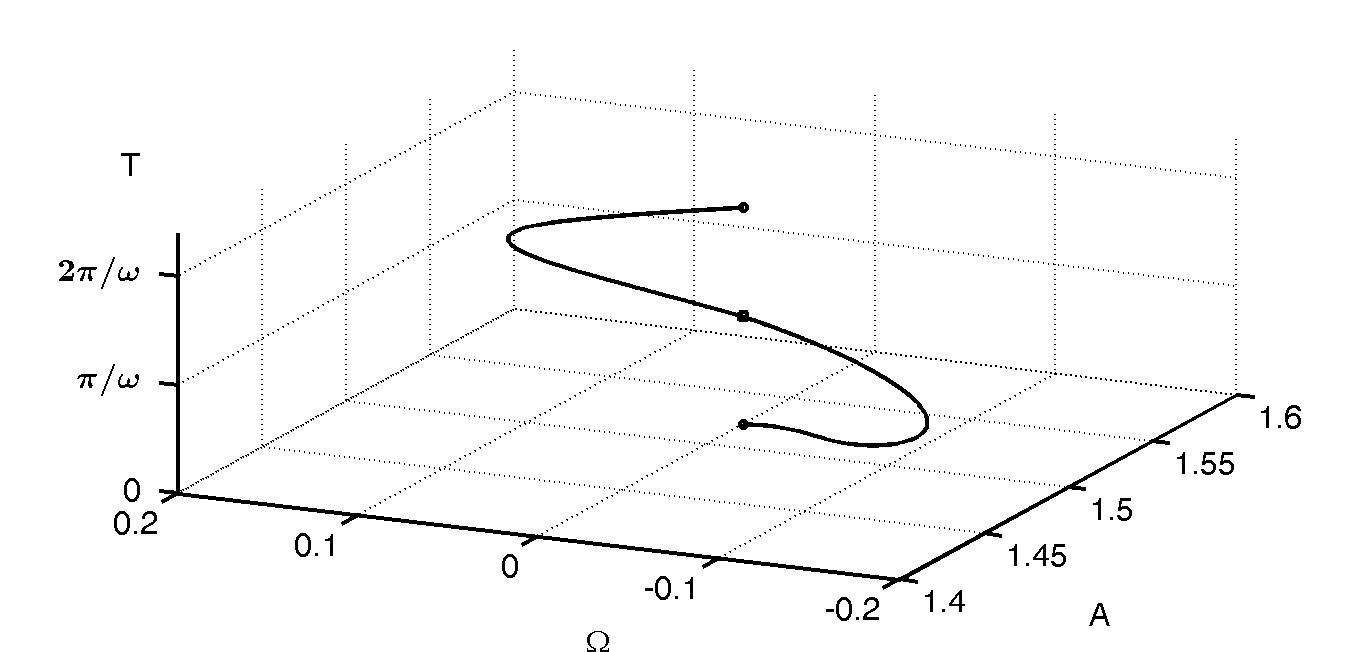}}
\caption{Optimal path for $\omega=1.5$.  Fixed points along path are indicated by circles (stable fixed points) or squares (saddles).}
\label{f:omega5}
\end{figure}

\begin{figure}[htbp]
\centerline{%
\includegraphics[width=\columnwidth]{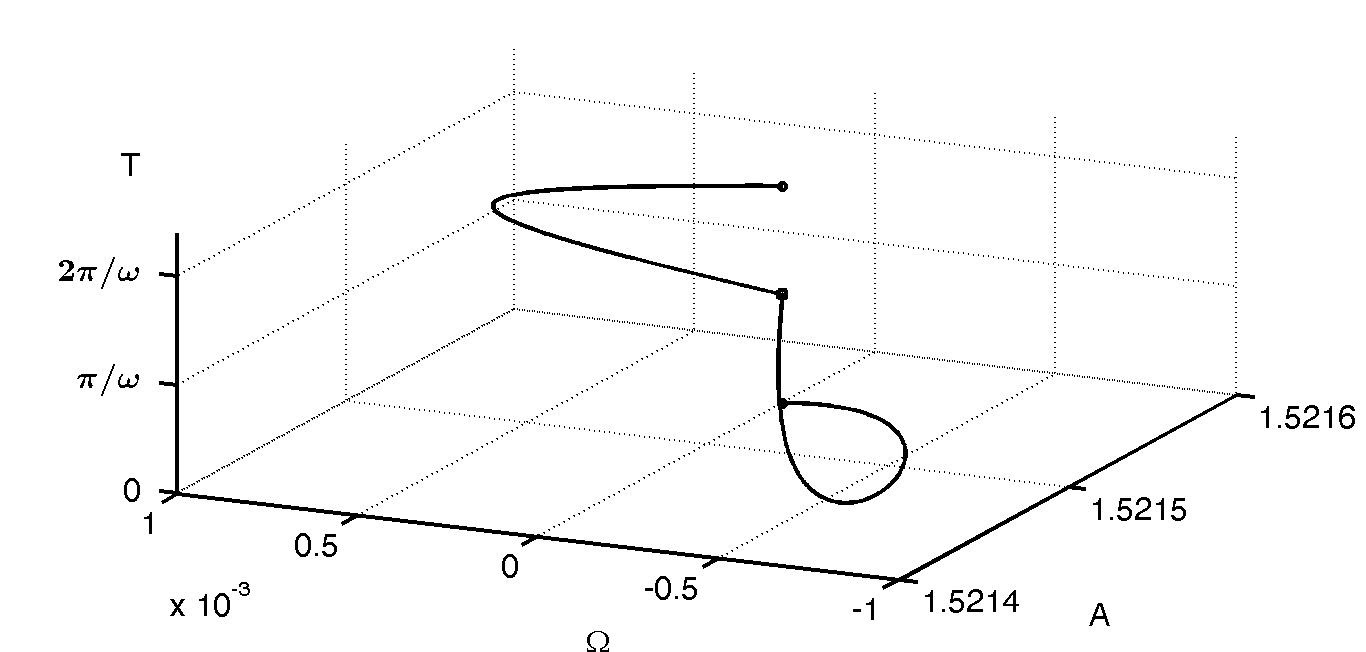}}
\caption{Optimal path for $\omega=10$.  Fixed points along path are indicated by circles (stable fixed points) or squares (saddles).}
\label{f:omega10}
\end{figure}

Figures~\ref{f:omega1} through~\ref{f:omega10} show the optimal paths obtained for $\omega=1$, $\omega=5$ and $\omega=10$, respectively.  Each path is drawn from its start at the stable fixed point $(A_0\; 0\; 0)^T$ to the nearest stable fixed point $(A_0\; 0\; 2)^T$, passing through the saddle at $(A_0\; 0\; 1)^T$.  The arrows in Fig.~\ref{f:tradeoff} identify the values of $\omega$ used for Figs.~\ref{f:omega1} through~\ref{f:omega10}.  It is evident from the axis scalings of the three plots that the $\omega=1.5$ optimal path undergoes the most significant deformation in amplitude $A$ and frequency $\Omega$ as it leaves the basin of the stable fixed point.  This suggests that it is better able to exploit the deterministic dynamics of Eqn.~\ref{e:SODE} to effect an exit than the $\omega=0.1$ or $\omega=10$ paths, leading to a lower barrier to cross for transitions between stable fixed points.  The two extreme values of $\omega=0.1$ and $\omega=10$ have paths that can also be explained by recalling that the restorative potential term $\cos\omega t$ in Eqn.~\ref{e:NLS_MLL} has a local curvature that is proportional to $\omega^2$ and a separation between fixed points that is inversely proportional to $\omega$.  Thus, at either extreme, the optimal exit path is one that simply translates away from the original fixed point without significantly changing $A$ or $\Omega$ en route to exit.  This is also reflected in the corresponding control term in Eqn.~\ref{e:SODE_RHS}, which vanishes as $\omega\rightarrow 0$ and as $\omega\rightarrow\infty$.

\section{Discussion}
\label{s:discussion}

In principle, optimizing the physical parameters relevant to a mode-locked laser's operation should take into account the entire invariant measure of Eqn.~\ref{e:SODE}.  Here, we have highlighted two aspects of that measure that are readily quantifiable to illustrate the trade-offs involved.

As mentioned above, Fig.~\ref{f:tradeoff} plots the exponential scaling factor in the slip rate but not the slip rate itself, which requires a normalization constant computed over all trajectories significantly contributing to the exit probability.  Such a computation typically requires rare-event sampling, which may be assisted through the use of the optimal paths to exit computed here~\cite{dupuis_stochastic_1987}.  This work, as well as the extension of these methods to more realistic laser feedback models incorporating delay, are the subject of ongoing research.

\section{Acknowledgments}

This material is based upon work supported by the National Science Foundation under Grant No.\ 1109278.

%

\end{document}